\documentclass{aa}

\usepackage{graphics}
\newcommand{\xbm}{\begin{math}}      \newcommand{\xem}{\end{math}}
\newcommand{\xbe}{\begin{enumerate}} \newcommand{\xee}{\end{enumerate}}
\newcommand{\xbi}{\begin{itemize}}   \newcommand{\xei}{\end{itemize}}
\newcommand{\xbeq}{\begin{eqnarray}} \newcommand{\xeeq}{\end{eqnarray}}
\def\be{\begin{equation}}
\def\ee{\end{equation}}
\def\be{\begin{equation}}
\def\ee{\end{equation}}
\def\bea{\begin{eqnarray}}
\def\eea{\end{eqnarray}}

\begin{document}

\title{Multiple scattering of polarized radiation
  by non-spherical grains: first results}
\titlerunning{Multiple scattering of polarized radiation
  by non-spherical grains}

\author{
  S. Wolf            \inst{1},
  N.V. Voshchinnikov \inst{2}, and
  Th. Henning        \inst{3}
  }
\authorrunning{Wolf et al.}

  \offprints{N.V. Voshchinnikov, nvv@astro. \protect\linebreak
  spbu.ru}

  \institute{
   \inst{1}
    Thuringian State Observatory Tautenburg,
    Sternwarte 5, D-07778 Tautenburg, Germany
    \\
    \inst{2}
    Sobolev Astronomical Institute, St.~Petersburg University,
    Universitetskii prosp. 28, 198504 Stary Peterhof-St.~Petersburg, Russia
    and  Isaac Newton Institute of Chile, St.~Petersburg Branch
    \\
    \inst{3}
    Astrophysical Institute and University Observatory (AIU),
    Friedrich Schiller University, Schillerg\"a\ss chen 3, D-07745 Jena, Germany
    }

  \date{Received $<$date$>$ / Accepted $<$date$>$}

  \abstract{
    We present the first {\rm numerical radiative transfer simulation of
    multiple light scattering} in dust configurations
    containing aligned non-spherical (spheroidal) dust grains.
{\rm     Such models are especially important if one wants
    to explain the circular polarization of light, observed in a variety of astronomical
    objects.}
    The optical properties of the spheroidal grains are calculated
    using the method of separation of variables developed by Voshchinnikov
    \& Farafonov~(\cite{vf93}).
    The radiative transfer problem is solved on the basis of the Monte Carlo
    method. Test simulations, confirming the correct numerical
    implementation of the scattering mechanism, are presented.
        As a first application, we investigate the
    linear and circular polarization of light coming from
    a spherical circumstellar shell. This shell contains perfectly aligned
    prolate or oblate spheroidal grains.
    We investigate the dependence of the results on the grain parameters
    (equivolume radius, aspect ratio) and
    the shell parameters (inner/outer radius, optical thickness).
    The most remarkable features of the simulated
    linear polarization maps
    are so-called polarization null points where the reversal of
    polarization occurs. They appear in the case when the grain alignment
    axis is perpendicular to the line of sight.
{\rm     The position of these points
    may be used for the estimation of grain shape and geometrical structure
    of the shell. The origin of null points lies in the
    physics of light scattering by non-spherical particles
    and is not related to the cancellation of polarization
    as  was discussed in previous models.}
    The maps of circular polarization have a sector-like
    structure with maxima at the ends of lines inclined
    to the grain alignment axis by $\pm 45\degr$.
      }

\maketitle

\section{Introduction}\label{intro}

It is now well established that the polarization of optical
{\rm and near-infrared radiation} from young and evolved stellar objects,
reflection nebulae, and active galactic nuclei is mainly caused by dust grains.
In many cases the observed polarization can be satisfactorily interpreted
by light scattering on spherical grains.
In particular, the orientation of the polarization vectors is often
used for {\rm the investigation of the
dust distribution and for identifying the location
of the embedded illuminating source(s).}

However, even in the case of  simple objects like ordinary
reflection nebulae, deviations of polarization vectors from the direction
perpendicular to a star were discovered more than three decades ago
(Elvius \& Hall~\cite{eh67}).
These deviations are displayed in the outer filamentary parts of the Merope
nebula {\rm and may be easily explained by light scattering on non-spherical grains
aligned by a magnetic field  along the filaments.}
Non-centrosymmetric polarization patterns have been observed
in bipolar and cometary nebulae (Scarrott et al.~\cite{sdw89}),
young stellar objects (Hajjar \& Bastien~\cite{hb96}),
evolved stars (Kastner \& Weintraub~\cite{kw96}),
and are also clearly seen in polarization maps of comets (Dollfus \&
Suchail~\cite{ds87}).
{\rm They may be attributed to non-spherical grains, although they could
be caused by multiple scattering on spherical particles as well.}
Another effect which may be related to the light scattering
by non-spherical grains is
the wavelength dependence of the positional angle of polarization
observed in red giants, AGB stars, and bipolar  reflection 
nebulae (see, e.g., Johnson \& Jones~\cite{jj91}). The variations
from blue to red may reach $20\degr - 60\degr$
and it is very difficult or even impossible to interpret this behaviour using
spherical grains only.

Recently, very high degrees of circular polarization 
of scattered light in the Orion molecular cloud
were measured by Chrysostomou et al.~(\cite{cgm00}).
The authors suggest that the circular polarization is produced by aligned 
non-spherical grains. {\rm Multiple scattering of radiation by spherical
or randomly oriented non-spherical grains results in a much smaller 
circular polarization degree than observed.}
In addition, the interstellar polarization
and polarized thermal emission phenomena prove
that non-spherical grains exist in the interstellar medium.
These effects arise because of dichroic extinction/emission of radiation
by aligned non-spherical grains and were modeled with
spheroidal grains (Voshchinnikov~\cite{v90}; Kim \& Martin~\cite{km95};
Onaka~\cite{o00}).

Up to now, spheroidal grains have been used for the interpretation of scattered
radiation only in the case of single scattering
(Voshchinnikov~\cite{v98}; Gledhill \& McCall~\cite{gm00}).
{\rm The numerical simulation of the polarized radiation transfer through a medium
with non-spherical grains, including multiple scattering, is extremely difficult.
In the case of spherical grains nearly all previous simulations are based
on Monte Carlo simulations.} This method was used
for the interpretation of polarimetric observations of various stars
(see, e.g., Daniel~\cite{d80}; Voshchinnikov \& Karjukin~\cite{vk94})
as well as the production of polarimetric maps of different
extended objects (e.g., Bastien \& M\'enard~\cite{bm88};
Fischer et al.~\cite{fhy94}; Wolf et al.~\cite{wfp98}).
The most recent, major achievement of the Monte Carlo technique 
--- in respect to the solution to the radiative transfer (RT) problem ---
is the development of RT codes that allow to calculate the spectral energy
distribution of multi-dimensional dust configurations self-consistently
(Wolf et al.~\cite{whs99}; Wolf \& Henning~\cite{wh00}).

{\rm The main aim of this paper is to provide the first
radiative transfer simulations including scattering,
extinction,  absorption, and re-emission of radiation
by aligned spheroidal dust grains.
Our formalism will be applied to several simple model configurations.
Because of the complexity of the problem, we decided to
consider the separate mechanisms step by step.
In this paper,  the basic theory is presented
and the main numerical features
of the simulation are described (Sect.~\ref{basic-all}).
Furthermore, we consider multiple scattering of light
by spheroidal grains and the resulting linear and circular polarization
(Sect.~\ref{test} and \ref{appl}).}

\section{Basics}\label{basic-all}

\subsection{Dust grain parameters}

Let us suppose that the dust grains are prolate or oblate homogeneous
spheroids with an aspect ratio $a/b$, where the quantities $a$ and $b$
are the major and minor semiaxes of a spheroid, respectively.
The particle size is characterized by the radius $r_{\rm V}$
of a sphere with the same volume as that of the spheroid
(equivolume radius). The major semiaxis $a$ of the spheroid 
is connected with $r_{\rm V}$ by
\be
a = r_{\rm V} \cdot \left(\frac{a}{b}\right)^{2/3}
\ee
for prolate spheroids and
\be
a = r_{\rm V} \cdot \left(\frac{a}{b}\right)^{1/3}
\ee
for oblate spheroids.
The optical properties of a particle also depend on
the complex refractive index of the grain material  $m_\lambda$ and
the angle $\alpha$
between the rotation axis of the spheroid and the wave-vector direction
(direction of the incident radiation; $0\degr \leq \alpha \leq 90\degr$,
see Fig.~\ref{scetch}).

\subsection{Radiative transfer concept}\label{mcm}

{\rm In this section we outline the main ideas of our solution
of the RT problem taking into account multiple scattering and dichroic absorption,
and reemission by spheroidal grains.
Our solution is based on the Monte Carlo method which is described
only briefly here. For a more general overview about
RT based on the Monte Carlo method, we refer to
the papers of Wolf et al.~(\cite{whs99}) and Wolf \& Henning~(\cite{wh00})
(see also Fischer et al.~\cite{fhy94}).

In our solution to the RT problem, the radiation energy
is partitioned into so-called weighted photons
(in the following we will use the term ``photon'').
The energy, intensity, and polarization of a weighted
photon are described by its wavelength $\lambda$ and the Stokes vector
$\hat{I} = (I,Q,U,V)^{\rm T}$. The transformation of the Stokes vector by
the $i$th scattering can be described by a $4\times4$ (M\"uller) matrix $\hat{F}$:
\begin{equation}
  \left(
    \begin{array}{c}
      I\\Q\\U\\V
    \end{array}
  \right)_i
  \propto
  \hat{F}
  \left(
    \begin{array}{c}
      I\\Q\\U\\V
    \end{array}
  \right)_{i-1}\ .
\end{equation}
}

The main features distinguishing the light scattering by aligned non-spherical
particles and spherical particles are:
\begin{enumerate}
\item Azimuthal dependence of the scattered radiation;
\item Linear polarization in the forward and
      backward directions;
\item Deviation of the positional angle of linear polarization
      after first scattering
      from the direction perpendicular to the illuminating source and
\item Circular polarization after first scattering.
\end{enumerate}
These features result from the fact that in the general case
of non-spherical particles --- in contrast to spherical grains ---
all 16 elements of the scattering matrix
are non-zero (see Voshchinnikov \& Farafonov~\cite{vf93}).

{\rm The spatial distribution of the illuminating star(s) and the dust grains
is defined inside a convex model space.
The free path length $l$ of a photon (distance between the point of emission and the first
point of scattering/absorption or two points of scattering/absorption) is determined
by the dust density distribution $n_{\rm d}(\vec{R}_{i})$ at the coordinate $\vec{R}_{i}$
and the extinction cross-section for the transmitted radiation
$\tilde{C}_{\rm ext}(\vec{R}_{i})$ }
\be \label{freepath}
\tilde{\tau}_{\rm ext} = \sum^{i_{\rm end}}_{i=1}
n_{\rm d}(\vec{R}_{i}) \cdot
\tilde{C}_{\rm ext}(\vec{R}_{i}) \cdot 
\Delta l_i \, .
\ee
{\rm The quantity $i_{\rm end}$  represents the number of integration steps
along the free path length,
and the quantity $\Delta l_i$ is the corresponding geometrical step width.}
The free mean path length
\be
l = \sum^{i_{\rm end}}_{i=1}
\Delta l_i \,
\ee
{\rm can be derived from the optical depth $\tilde{\tau}_{\rm ext}$
as follows
\be
\tilde{\tau}_{\rm ext} = -\ln(1-\zeta)\ .
\ee
Here, $\zeta$ is a random number uniformly distributed on the interval
$[0,1]$ (see Wolf et al.~\cite{whs99} for details).}

The extinction cross-section $\tilde{C}_{\rm ext}$ is calculated 
taking into account the state of polarization of the incident radiation 
(see, e.g., Martin~\cite{m74}):
\bea \label{avc}
\tilde{C}_{\rm ext} = \frac{1}{2}(C_{\rm ext}^{\rm TM}+C_{\rm ext}^{\rm TE}) +
\frac{1}{2}(C_{\rm ext}^{\rm TM}-C_{\rm ext}^{\rm TE}) \cos 2\Psi \frac{Q_0}{I_0} +
\nonumber \\
\frac{1}{2}(C_{\rm ext}^{\rm TM}-C_{\rm ext}^{\rm TE}) \sin 2\Psi \frac{U_0}{I_0} \,.
\eea
Here,  $\Psi$ is the angle between the particle frame and the laboratory frame.
$(I_0, Q_0, U_0, V_0)$ is the Stokes vector of the incident light.
The extinction cross-sections $C_{\rm ext}^{\rm TE,TM}$ are connected with
the efficiency factors $Q_{\rm ext}^{\rm TE,TM}$ by
\be
C_{\rm ext}^{\rm TE, TM} = G \cdot Q_{\rm ext}^{\rm TE,TM} \, ,
\ee
where the quantity $G$ is the geometrical cross-section of the dust grain 
(the area of the particle shadow). The superscripts TM and TE 
are related to the two cases of the polarization of the incident radiation 
(TM and TE modes).
The geometrical cross-section $G$ can be derived as follows
\be
G(\alpha) = \pi r_{\rm V}^2 \cdot \left (\frac{a}{b} \right ) ^{-2/3}
            \cdot \left[\left (\frac{a}{b} \right )^{2}\sin^2\alpha
            + \cos^2\alpha\right]^{1/2}
\label{Gp}
\ee
for a prolate spheroid and
\be
G(\alpha) = \pi r_{\rm V}^2 \cdot \left (\frac{a}{b} \right ) ^{2/3}
            \cdot \left[\left (\frac{a}{b} \right )^{-2}\sin^2\alpha
            + \cos^2\alpha \right]^{1/2}\,.
\label{Go}
\ee
for an oblate spheroid.
The efficiency factors  $Q_{\rm ext}^{\rm TM,TE}$ are calculated from
the solution to the light scattering problem for spheroids
obtained by the method of separation of variables
(see Voshchinnikov \& Farafonov~\cite{vf93} for details).

We would like to note  that the extinction cross-section
$\tilde{C}_{\rm ext}$ (see Eq.~(\ref{avc}))
has to be determined at each integration step along the photon path
(see Eq.~(\ref{freepath})).
In the case of the RT in a medium with aligned non-spherical grains,
the step width along the photon free path is determined
not only by the scale of the density structure (as for spherical particles)
but also by the degree of spatial variations of the grain alignment.
Owing to this and the more complex treatment of the scattering process compared
with spherical grains, the total computation run-time is by an order of magnitude 
larger for aligned spheroids than for spheres.

{\rm Finally, depending on the model geometry and the dust density distribution,
the photon leaves the model space after a certain number of interactions
with dust grains (scattering, absorption).
Then, it will be ``observed'' by array-detector-like planes arranged
around the model space. The Stokes vector has to be projected
onto the respective plane. For a detailed description of the projection
formalism we refer to Fischer et al.~(\cite{fhy94}).}

\subsection{Scattering }\label{scapro}

\begin{figure} 
  \resizebox{\hsize}{!}{\includegraphics{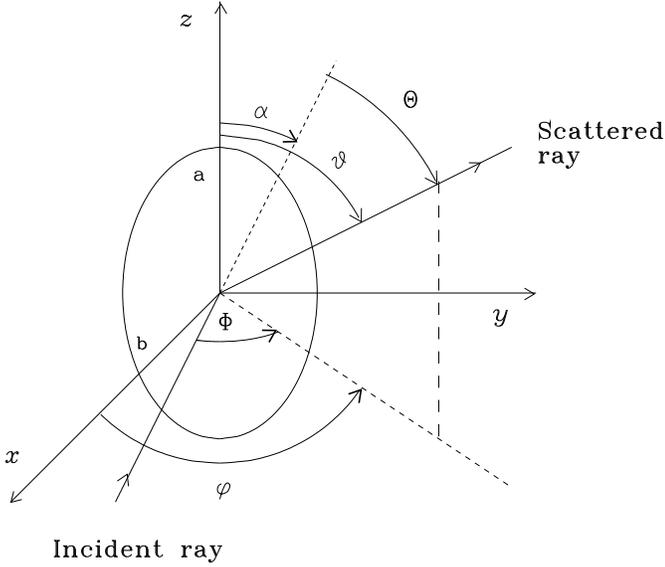}}
  \caption[]{Scattering geometry for prolate homogeneous spheroid.
The origin of the coordinate system is at the centre
of the spheroid while the $z$-axis coincides with its axis of revolution.
The angle of incidence $\alpha$ is the angle in the $x-z$-plane
between the direction of incidence and the $z$-axis.
The scattered field in the far-field zone is represented in a spherical
coordinate system by the angles $\theta, \varphi$
in the reference system related to the spheroid's rotation axis
(particle frame) or by the angles $\Theta, \Phi$ in the reference system 
related to the scattering plane (laboratory frame).}
  \label{scetch}
\end{figure}

The solution to the problem of light scattering by spheroidal particles
is given in a particle reference system related to the
rotation axis of a spheroid (in spherical coordinates $\theta, \varphi$), while
the RT is considered in the laboratory reference
system (in spherical coordinates $\Theta, \Phi$; {\rm see Fig.~\ref{scetch}).
The probability of a photon of being scattered into the direction
($\Theta,\Phi$) is proportional to the value of the first element
of the scattering matrix $F_{11}(\alpha, \Theta,\Phi)$.}
We transformed the scattering matrix
from the particle to the laboratory frame.
Then,  the scattering matrix is calculated  using the standard expressions
(Bohren \& Huffman~\cite{bh83}).
In order to check the transition from the particle
system to the laboratory one,
the transformation was also done
for light scattering by spheroidal particles in the quasistatic approximation
where the solution can be written in both coordinate systems
(see Farafonov et al. \cite{fip01} for details).

The elements of the scattering matrix were normalized
using the standard procedure (Bohren \& Huffman~\cite{bh83})
\bea\label{normalize}
 \frac{2}{k^2 (C_{\rm sca}^{\rm TM} + C_{\rm sca}^{\rm TE})}
 \int_0^{2 \pi} \! \int_0^{\pi}
F_{11}(\alpha, \Theta, \Phi)  
 \sin \Theta {\rm d}\Theta {\rm d}\Phi  = 1 ,
\eea
where $C_{\rm sca}$ is the scattering cross-section and
$k = 2 \pi /\lambda$ the wavenumber.
The correctness of these computations was verified
using the relationships for  matrices, describing scattering by small
particles (see Hovenier \& van der Mee~\cite{hv00} and references
therein).

\begin{figure} 
  \resizebox{\hsize}{!}{\includegraphics{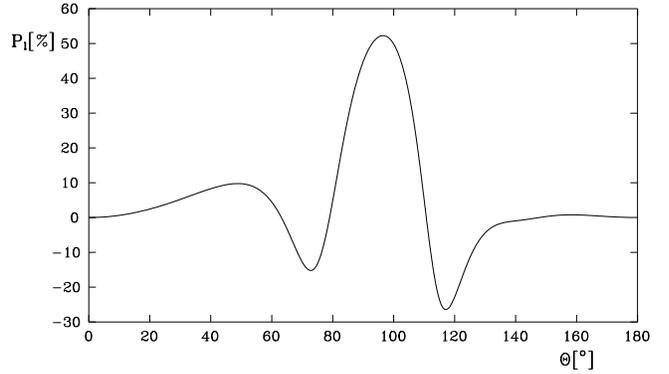}}
  \caption[]{
    The linear polarization of radiation $P_{\rm l}(\Theta)$ after the first
    scattering by a prolate spheroidal particle
    with
    refractive index $m=1.7+0.03i$
    (typical of astronomical silicate at optical wavelengths),
    equivolume radius $r_{\rm V} = 0.20\,{\rm \mu}$m,
    aspect ratio $a/b=4$ and
    the incidence angle $\alpha=0\degr$.
    The wavelength of incident radiation is $\lambda = 0.628 \,\mu$m.
    To avoid skipping of local minima/maxima a step size
    smaller than $5\degr$ for $\Theta$ is required.
    }
  \label{plex}
\end{figure}

\begin{figure} 
  \resizebox{\hsize}{!}{\includegraphics{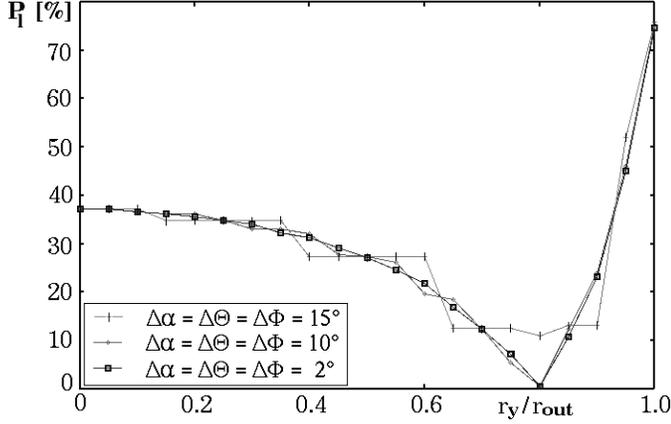}}
  \caption[]{
    The degree of linear polarization in the midplane of the polarization
    map shown in Fig.\,\ref{matmod}[A] in dependence of the projected
    distance.
    Three different discretizations for the angles
    $\alpha, \Theta,$ and $\Phi$ were applied
    ($\Delta\alpha = \Delta\Theta = \Delta\Phi = 2\degr, 10\degr$, and $15\degr$).
    Only the smallest step width gives a sufficient approximation
    to the analytical solution presented in Fig.\,\ref{matmod} [B1]
    (see Sect.~\ref{scapro} for discussion).
    }
  \label{discret}
\end{figure}

{\rm The calculation of the elements of the scattering matrix
$\hat{F}(m,r_{\rm V},{a}/{b},\alpha,\Theta,\Phi)$
and efficiency factors $Q_{\rm sca,ext}^{\rm TM,TE}(m,r_{\rm V},{a}/{b},\alpha)$
for a single particle with the fixed equivolume radius $r_{V}$, 
the aspect ratio $a/b$, 
and the refractive index $m$ requires several minutes of computer 
time\footnote{A Pentium\,II 450\,MHz processor was used for our calculations.}.
These quantities were precalculated before starting
the RT modelling.
In order to guarantee that
the main features (local minima/maxima) of the scattering matrix elements
  (and/or derived quantities, for instance, the polarization
  of singly scattered light)  are present in the discrete distributions
  as a function of $\alpha$, $\Theta$, and $\Phi$
(see Figs.\,\ref{plex} and \ref{discret}), we use step widths of
$\Delta\alpha = \Delta\Theta = \Delta\Phi = 2\degr$ in our simulations
(see Fig.~\ref{discret}).
}

\subsection{Absorption and Reemission}\label{abspro}

The absorption is described by the albedo $\Lambda$ of the particle which
is defined as the ratio of the scattering and extinction cross sections.
It is equal to
\begin{equation}\label{alb}
\Lambda^{\rm TM,TE}(m,r_{\rm V},{a}/{b},\alpha) =
\frac{C_{\rm sca}^{\rm TM,TE}(m,r_{\rm V},{a}/{b},\alpha)}
{C_{\rm ext}^{\rm TM,TE}(m,r_{\rm V},{a}/{b},\alpha)}
\end{equation}
for totally polarized light and
\begin{equation}\label{alb0}
\Lambda^{(0)}(m,r_{\rm V},{a}/{b},\alpha) =
\frac{C_{\rm sca}^{\rm TM}+
C_{\rm sca}^{\rm TE}}
{C_{\rm ext}^{\rm TM}+
C_{\rm ext}^{\rm TE}}
\end{equation}
for non-polarized light.

While in the case of absorption by spherical particles, the Stokes vector
$\hat{I}$ can be multiplied by a scalar value of the albedo,
for spheroidal particles the following $4 \times 4$ matrix is needed
(in the particle frame):
\begin{equation}\label{albmat}
\hat{\Lambda}_{\rm part.}  =
\left(
\begin{array}{cccc}
\tilde{l}_{\rm 1} & \tilde{l}_{\rm 2} &       0    & 0\\
\tilde{l}_{\rm 2} & \tilde{l}_{\rm 1} &       0    & 0\\
0         &      0    &  \tilde{l}_{\rm 1} & 0\\
0         &      0    &       0    & \tilde{l}_{\rm 1}\\
\end{array}
\right)_{\rm part.}   \,.
\end{equation}
Here, the quantities $\tilde{l}_{\rm 1}$ and $\tilde{l}_{\rm 2}$ are defined as
\begin{eqnarray}\label{albdef}
\tilde{l}_{\rm 1} = \Lambda =
\frac{\Lambda^{\rm TM}}{1+{C_{\rm ext}^{\rm TM}}/{C_{\rm ext}^{\rm TE}}}+
\frac{\Lambda^{\rm TE}}{1+{C_{\rm ext}^{\rm TE}}/{C_{\rm ext}^{\rm TM}}},
 \nonumber \\
\tilde{l}_{\rm 2} =
\frac{\Lambda^{\rm TM}}{1+{C_{\rm ext}^{\rm TM}}/{C_{\rm ext}^{\rm TE}}}-
\frac{\Lambda^{\rm TE}}{1+{C_{\rm ext}^{\rm TE}}/{C_{\rm ext}^{\rm TM}}}\,.
\end{eqnarray}

Before applying the albedo matrix $\hat{\Lambda}_{\rm part.}$,
it has to be rotated into the particle frame using the rotation matrix $\hat{R}$
\begin{equation}\label{rotmat}
\hat{R} = 
\left(
\begin{array}{cccc}
1 &      0      &       0      & 0\\
0 & \cos2\Psi   &  \sin2\Psi   & 0\\
0 & -\sin2\Psi  &  \cos2\Psi   & 0\\
0 &      0      &       0      & 1\\
\end{array}
\right).
\end{equation}
The absorption process can then be described by the expression
\begin{equation}\label{salto}
  \hat{I}_{\rm after} =   
  \hat{R}^{-1}\, \hat{\Lambda}_{\rm part.}\, \hat{R}\, \hat{I}_{\rm before}\,,
\end{equation}
whereby $\hat{I}_{\rm before}$ ($\hat{I}_{\rm after}$) is the Stokes vector
of the radiation before (after) absorption.

{\rm In the case of (re)emission of radiation by a spheroidal
particle, the Stokes vector of the photon can be written as follows
(particle frame):
\be
\left( \begin{array}{c} I_0 \\
                        Q_0 \\
                        U_0 \\
                        V_0 \end{array} \right)_{\rm part.}
\propto
\left( \begin{array}{c} (C_{\rm abs}^{\rm TM}+C_{\rm abs}^{\rm TE})/2 \\
                        -(C_{\rm abs}^{\rm TM}-C_{\rm abs}^{\rm TE})/2 \\
                        0 \\
                        0 \end{array} \right)_{\rm part.}
                        B_{\lambda}(T_{\rm d})  \, .
\ee
Here, $C_{\rm abs}$ is the absorption cross-section,
$B_{\lambda}(T_{\rm d})$ the Planck function and
$T_{\rm d}$ the grain temperature.
}

\section{Test simulations}\label{test}

The best way to test any RT code is the comparison of 
its results with those obtained with another RT code/technique or an analytical solution.
Because the RT code presented here is the {\sl first} that provides the treatment
of multiple scattering of polarized light by aligned non-spherical
particles, a lack of suitable test cases for comparison is evident.
Nevertheless, in the following we present test calculations
which confirm the correct implementation of the new scattering
mechanism into our RT code.

For test simulations we used particles with the following parameters:
\begin{itemize}
\item shape of the grains:              oblate,
\item refractive index:                 $m=1.7+0.03i$,
\item equivolume radius:                $r_{\rm V}=0.1\,\mu$m,
\item aspect ratio:                     $a/b=2.0$,
\item wavelength of incident radiation: $\lambda = 0.628 \,\mu$m.
\end{itemize}
Then the  grain  size parameter is $x_{\rm V}={2\pi r_{\rm V}}/{\lambda}=1$.

\subsection{Single scattering in a geometrically and optically thin hemisphere}\label{tesisc}

\begin{figure} 
  \resizebox{\hsize}{!}{\includegraphics{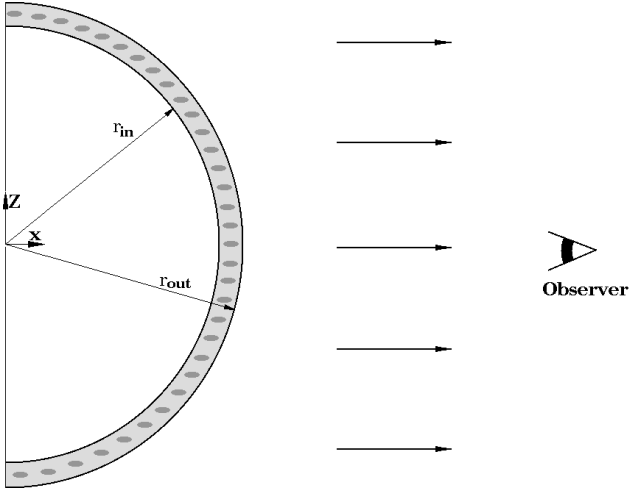}}
  \caption[]{Model configuration for comparison with the
    results of Matsumura \& Seki~(\cite{ms96}). It consists of a thin
    dust layer at the surface of a hemisphere. The ratio of
    the inner radius to the  outer one amounts to $r_{\rm in}/r_{\rm out}=0.99$.
    The shell contains oblate grains, whereby their rotation axes ($b$) are
    oriented parallel to the $z$-axis. The dark grey oval pattern symbolizes
    the aligned grains. The configuration is shown
    for $y=0$ ($xz$-plane,  the azimuthal angle $\Phi=180\degr$).
    An observer sees the radiation at the scattering angles
    $\Theta=0\degr-90\degr$.
    In this case the condition $\Theta + \alpha =90\degr$
    is satisfied in all points of the dust layer. }
  \label{modsisc}
\end{figure}

Firstly, we consider single scattering in a thin layer
of oblate spheroids at the surface of an hemisphere (see Fig.~\ref{modsisc}).
The ratio of the inner radius ($r_{\rm in}$) of the shell to the outer one
($r_{\rm out}$) amounts to  $r_{\rm in}/r_{\rm out} = 0.99$.
The non-rotating particles show a static picket fence orientation with their
rotation axes being aligned parallel to the $z$-axis of the global
coordinate system of the model space.
The density profile $n(r)$ in the shell follows a power law:
\begin{equation}\label{denprof}
n(r) \propto r^{-1}\,.
\end{equation}

The optical thickness of the shell at the wavelength of the simulated RT
($\lambda = 0.628 \,\mu$m) is 0.2 along the $z$-axis,
i.e. $\tau_{\rm z}(0.628 \,\mu{\rm m})=0.2$.
Because of the chosen grain alignment, $\tau_{\rm z}$ is proportional
to the extinction
cross-section  in the case of light incident parallel to the grain rotation axis,
$\tau_{\rm 0} = \tau_{\rm z} \propto C_{\rm ext} (0\degr)$.
In other directions, the optical thickness of the shell may be found as
$\tau_{\alpha} = \tau_{\rm z} C_{\rm ext} (\alpha)/C_{\rm ext} (0\degr)$
and, in particular, along the line of sight ($x$-axis) and the $y$-axis
we have
$\tau_{\rm x} = \tau_{\rm y} =\tau_{90\degr} =
\tau_{\rm z} C_{\rm ext} (90 \degr)/C_{\rm ext} (0\degr)$.
Using Eq.~\ref{avc} for non-polarized stellar radiation and the standard
model described above, we have
$\tau_{90\degr} \approx  0.35$ (prolate grains) and
$\tau_{90\degr} \approx  0.13$ (oblate grains), if
$\tau_{0}(0.628 \,\mu{\rm m})=0.2$. For more elongated or flattened
particles ($a/b=10$), these values become equal to
$\tau_{90\degr} \approx  0.68$  and 0.082 for
prolate and oblate grains, respectively.

\begin{figure*} 
  \resizebox{\hsize}{!}{\includegraphics{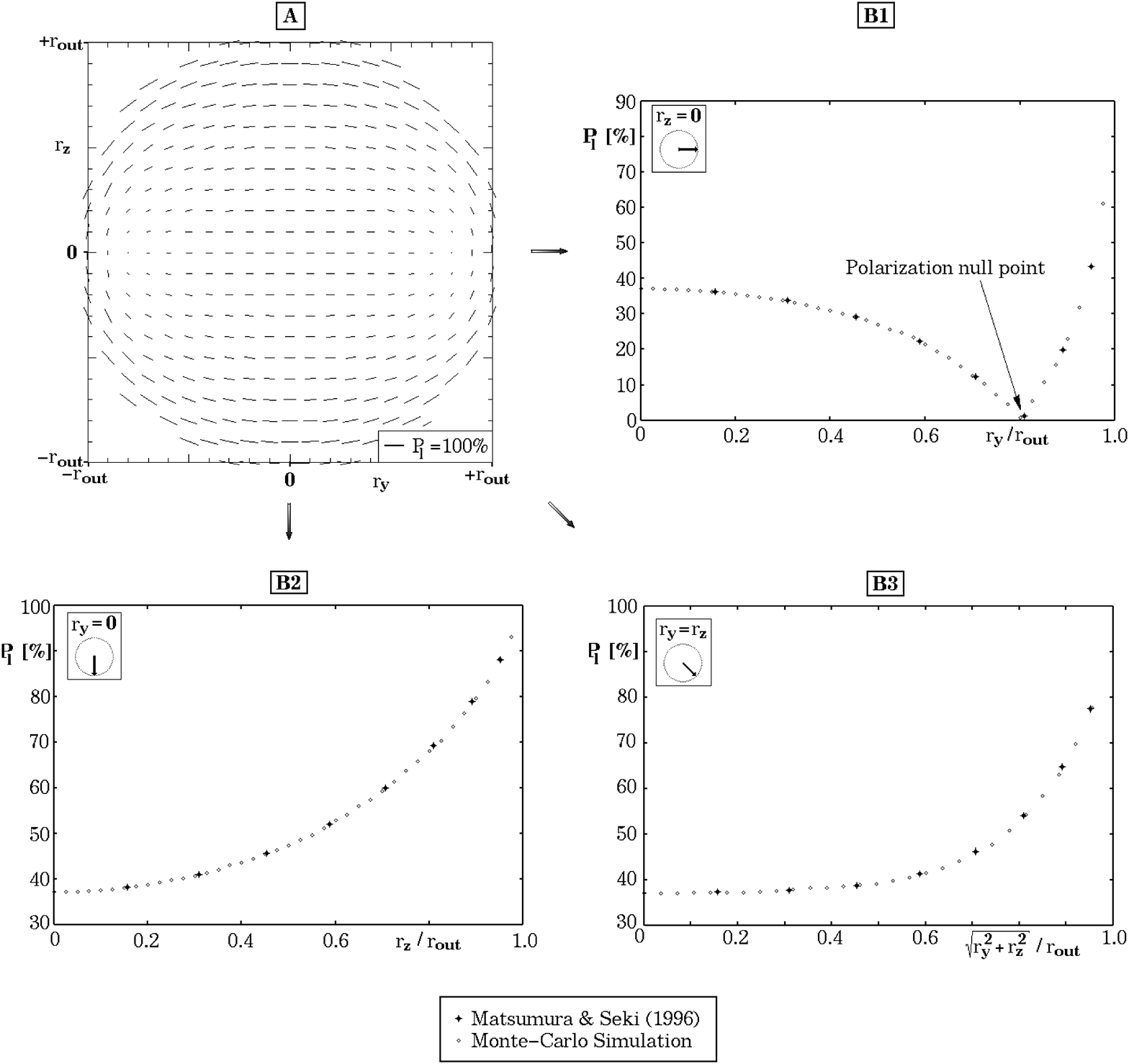}}
  \caption[]{
    The results of test calculations for the comparison with
    the data of Matsumura \& Seki~(\cite{ms96}).
   {\bf [A]}: Polarization map of the configuration shown in Fig.~\ref{modsisc}.
    Only single scattering is considered.
    {\bf [B1]}$\ldots${\bf [B3]}: The degree of linear polarization
    as a function of the projected distance from the centre
    along three different directions.
    The projected distance $r/r_{\rm out}$ is connected with
    the scattering angle as  $r/r_{\rm out}= \arcsin \Theta$.
    {\bf [B1]}: $P_{\rm l}$ in the midplane of polarization map;
    {\bf [B2]}: $P_{\rm l}$ perpendicular to the midplane;
    {\bf [B3]}: $P_{\rm l}$ along a line being inclined by
    $45\degr$  to the midplane.}
  \label{matmod}
\end{figure*}

In contrast to polarization maps obtained for model configurations
with the same geometry, but containing spherical dust grains,
the polarization pattern is {\em not} centro-symmetric
(see Fig.~\ref{matmod}[A]).
This is because of the azimuthal asymmetry of light scattering by aligned
non-spherical particles. The net polarization
in this case is not equal to zero.
We compare our results with those obtained by 
Matsumura \& Seki~(\cite{ms96})\footnote{The numerical
data in tabular form were provided by Matsumura~(\cite{m99}).}.
In Fig.~\ref{matmod} the comparison of the polarization degree
as a function of the projected distance from the centre along
three different lines in the polarization map is shown.
These graphs as well as the polarization pattern
(see for comparison Fig.~2 from  Matsumura \& Seki~\cite{ms96})
are in perfect agreement.

Qualitatively, the polarization pattern is quite similar to that of a
young stellar object surrounded by a spherical shell and an optically thick
disk seen edge-on. {\rm There the parallel polarization pattern
is due to light scattering on the disk surface while the
centro-symmetric polarization pattern is due to light scattering in the
circumstellar environment.
Corresponding RT simulations with spherical
dust grains have been performed
by M$\acute{\rm e}$nard~(\cite{m88}) and {\nobreak Fischer et al.~(\cite{fhy94})}.
The most interesting feature of the pattern is the presence of polarization null points
(see Figs.~\ref{matmod}[A] and \ref{matmod}[B1]). They have been observed 
in a number of objects (see discussion in Sect.~\ref{null}).
Sometimes, these null points are interpreted as the indicators for
the ``outer edge'' of a circumstellar disk
(transition from the optically thick disk to the optically thin shell;
see Fischer et al.~\cite{fhy94}).
Although this explanation is very plausible, the simulations presented here show
that uniformly aligned grains (possibly by an external magnetic field)
may produce a similar polarization pattern.
A detailed discussion of the polarization null points
as a function of the model parameters is given in Sect.~\ref{appl}.

}

\subsection{Multiple scattering}\label{temusc}

To test the RT with multiple scattering, we consider
model configurations for which the dust density distribution,
the dust grain orientation, and the irradiation characteristic
are centro-symmetric with respect to the observer's position. 
The resulting polarization pattern is therefore also centro-symmetric:
the polarization vectors are oriented perpendicular or parallel
to the radius vector of the configuration, in dependence of the refractive
index and the grain size parameter $x_{\rm V}$.
Thus, the integral (net) linear and circular polarization have to be zero.
Because this property of centro-symmetric configurations does not depend
on its optical thickness, it can be used as a test of the single
and multiple scattering case.

We considered the following dust configurations containing either
oblate or prolate grains:
{\rm \begin{enumerate}
\item Spherical dust shell with a point-like radiation source
      in the centre and
     particle rotation axes either parallel or perpendicular
      to the radius vector;
    \item Cylindrical dust shell with a point-like radiation source
      on its rotation axis and
 particle rotation axes either parallel or perpendicular
      to the radius vector which is oriented perpendicular to the rotation
      axis of the cylinder.
\end{enumerate}}
While for the first configuration the net polarization is equal to zero for
every position of the observer, for the second configuration this is the case
only if the observer's line of sight is oriented perpendicular
to one of the two cylinder footprints.
{\it In all cases we arrived at the expected results.}

\section{Results and discussion}\label{appl}

{\rm In Sect.~\ref{moda}, light scattering in
a geometrically thin hemisphere and sphere and a geometrically thick
spherical shell is described. This preliminary modelling provides
a ``smooth'' transition from the test models discussed
in Sect.~\ref{test} to more realistic dust configurations.
The basic modelling is presented in  Sect.~\ref{modb},
where a geometrically thick shell is considered and
the resulting polarization as a function of the main model
parameters is investigated.}

If one considers the RT in a medium consisting of spherical dust grains,
scattering is the only mechanism to produce or to modify the polarization of
the incident radiation. In the case of aligned spheroidal grains,
the processes of photon absorption and re-emission may also
fabricate or change polarization.
Because of the obvious complexity of the problem, we separated
the consideration of these three mechanisms of the polarization formation.
The aim of the simulations discussed below was to outline the basic
features of polarization arising from the {\em light scattering} by spheroidal grains.
The additional influence of absorption and re-emission
will be discussed in future publications.

For this reason, we apply the following approximations:
\begin{enumerate}
\item The grain albedo is assumed to be that of spherical grains
  with the equivolume radius $r_{\rm V}$ and is therefore a scalar.
  Then, the absorption process decreases the intensity of the incident
  radiation but does not modify the state of polarization.
\item We do not consider thermal re-emission of radiation.
\end{enumerate}

{\rm If not stated differently, for the simulation presented in this section
the dust grain and model parameters as described in Sect.~\ref{test}
have been used.
The observer's position is inclined by $i=90^{\rm o}$ to the $z$-axis
($\tau_{\rm z}(0.628 \,\mu{\rm m})=0.2$)
of the global coordinate system of the model space (see Fig.~\ref{modsisc}).
}
\subsection{Simple models}\label{moda}

\begin{figure} 
  \resizebox{\hsize}{!}{\includegraphics{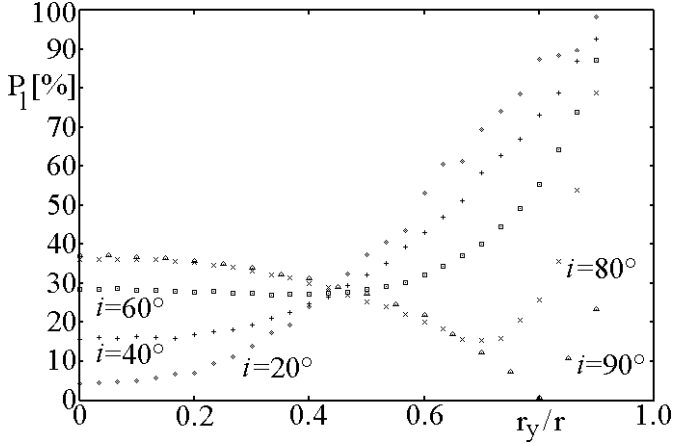}}
  \caption[]{
    The degree of linear polarization in dependence of the projected distance
    for tilted dust configurations.
    }
  \label{viewang}
\end{figure}
{\rm Let us consider again the geometrically thin hemispherical model.}
While the polarization pattern in Fig.~\ref{matmod} is shown only
for a position perpendicular to the $z$-axis ($i=90\degr$),
Fig.~\ref{viewang}
demonstrates how the behaviour of the degree of linear polarization
in the midplane of the polarization map appears for different inclinations
of the $z$-axis to the line of sight.
It turns out that the local polarization
minimum (the ``polarization null point''
or the reversal of polarization in the case $i=90^{\rm o}$)
are present at large inclination angles $i$ only.
It is important to note that the polarization pattern
(and therefore also the net polarization degree)
for aligned non-spherical particles
depends on the viewing angle even in the case of a one-dimensional density
distribution. In contrast to this, no dependence
of the polarization on the inclination angle
exists for spherical particles.

\begin{figure}
 \resizebox{\hsize}{!}{\includegraphics{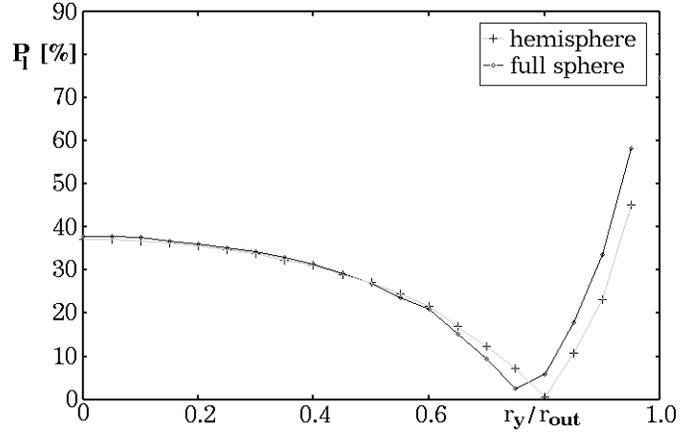}}
  \caption[]{
    The degree of linear polarization in dependence of the projected distance.
    Influence of the backward scattering is illustrated.
    }
  \label{backward}
\end{figure}
As a first step away from the very artificial configuration
described above, we consider the RT
in a thin spherical shell instead of the hemisphere shown
in Fig.~\ref{modsisc}. Then, backward scattering in the hemisphere opposite to
the observer's position (where $\Theta = 90\degr +\alpha$)
has to be taken into account.
The results are shown in Fig.~\ref{backward}.
The addition of the backscattered radiation shifts  the position of the
minimum of the linear polarization to a smaller projected radius.
\begin{figure*}[!htbp]
 \resizebox{\hsize}{!}{\includegraphics{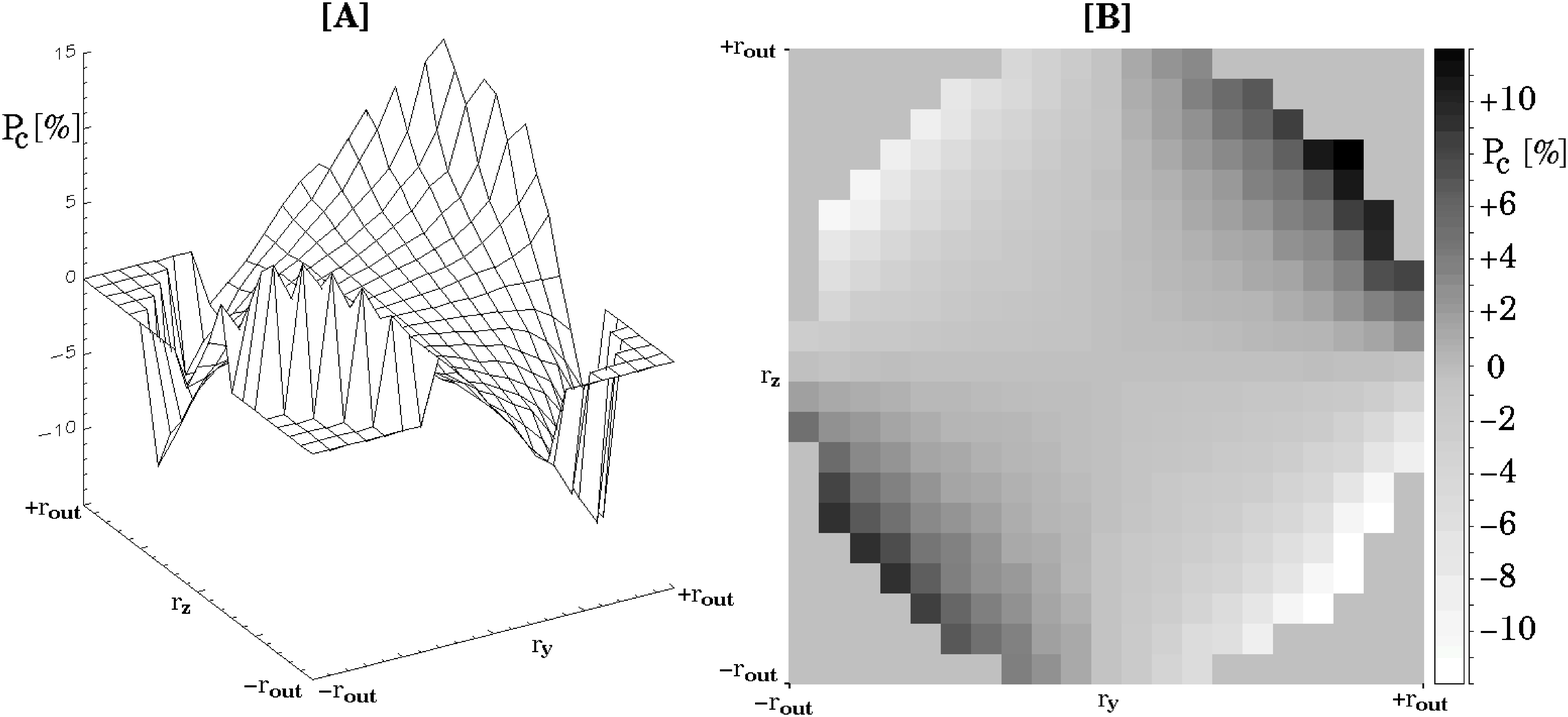}}
  \caption[]{
    {\bf [A]}
    The circular polarization of the scattered radiation
    arising from the configuration shown in Fig.~\ref{modsisc}.
    Deviations from the mirror symmetry result from the statistical noise
    of the Monte Carlo simulation.
    {\bf [B]}
    The same as grey plot to emphasize the symmetry of the spatial
    dependency of the circular polarization degree.
    }
  \label{circpol}
\end{figure*}

In Fig.~\ref{circpol}, the circular polarization
for this model of a geometrically and optically thin dust shell is shown.
The circular polarization is equal to zero in the midplane and along the line
through the centre perpendicular to the midplane.
The net circular polarization from the configuration is also
equal to zero.
A counterpart of the linear polarization null points
in the midplane could not be found.
The maximum circular polarization is  observed at the ends of lines inclined to
the $z$-axis by $\pm 45\degr$. These points correspond to the
light scattering by particles with values of
$ \alpha = 45\degr$, $\Theta = 90\degr$
and $\Phi = 90\degr \ (270\degr)$.

\begin{figure}
 \resizebox{\hsize}{!}{\includegraphics{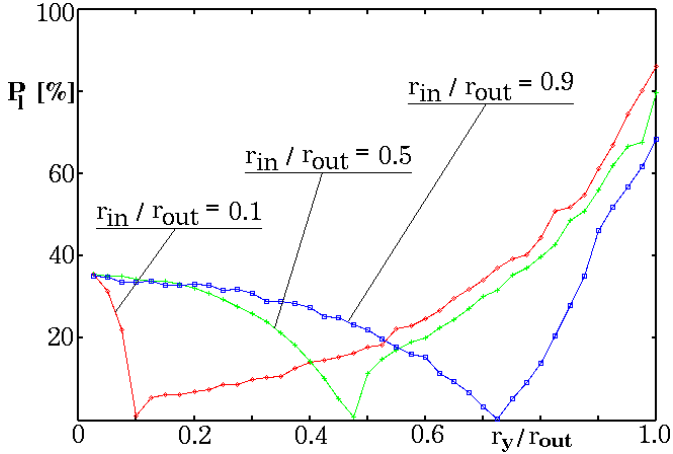}}
  \caption[]{
    The degree of linear polarization in dependence of the projected distance.
    The effect of the transition from a
    geometrically thin to a  geometrically
    thick spherical shell is illustrated.}
  \label{shell2sphere}
\end{figure}
{\rm It turned out that the pattern of the linear polarization strongly depends on
the inner radius of the spherical shell. This is demonstrated
in Fig.~\ref{shell2sphere}, where the degree of linear polarization
in the midplane is shown for three ratios of the inner to the outer
radius of the shell.
Decreasing this ratio (and therefore decreasing the relative size
of the inner cavity around the illuminating star), the characteristic polarization
null points shift towards the star. It has to be noted, that 
generally the polarization null point is not located
at the position of the projected inner radius.
The location, relative depth and width of the area where
the polarization changes the sign depends on the geometrical
thickness of the shell and  grain parameters (see Sect.~\ref{grpar}).
The power-law density profile used (see Eq.~(\ref{denprof}))
leads to a preferred photon scattering near a
narrow region at the inner boundary of the shell. The shift of the polarization
null point is therefore mainly caused by the change of the inner radius $r_{\rm in}$.
}

\subsection{Envelope models}\label{modb}

{\rm Now we will consider geometrically thick spherical shells.
To represent such models, the ratio of the inner radius of the shell to the outer 
radius is fixed at $r_{\rm in}/r_{\rm out}=0.5$.}
\subsubsection{Transition from optically thin to optically thick shell}\label{inoopdp}

\begin{figure}
 \resizebox{\hsize}{!}{\includegraphics{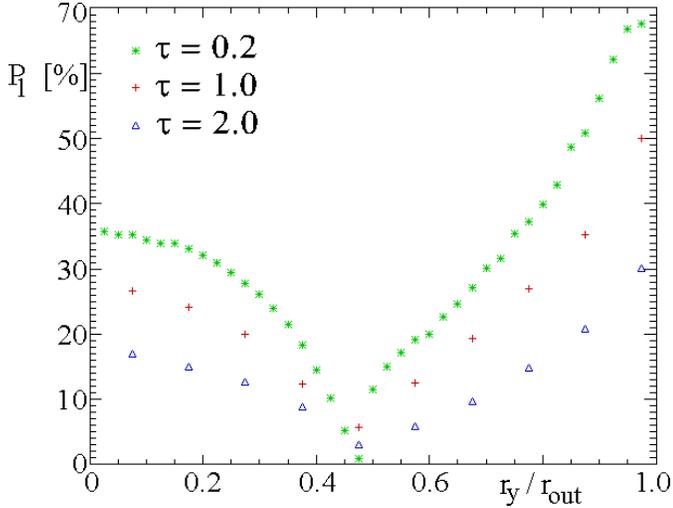}}
  \caption[]{
    The degree of linear polarization in dependence of the projected distance.
    The effect of the transition from an
    optically thin to an  optically
    thick spherical shell is illustrated.
    Note that at this Figure the resolution
    along the $y$-axis is much smaller than in comparison with other Figures.}
  \label{optdepth}
\end{figure}

In the case of a small optical thickness, the main contribution
to the resulting polarization is from first or second scattering events
(Voshchinnikov \& Karjukin~\cite{vk94}; Voshchinnikov et al.~\cite{vgk95}).
As one can see from Fig.~\ref{optdepth}, the polarization
$P_{\rm l}(r_{y}/r_{\rm out})$ in the midplane
decreases with increasing optical thickness. The same behaviour was found
for the polarization perpendicular to the midplane $P_{\rm l}(r_{z}/r_{\rm out})$.
In general, the position of the local polarization minimum
is not strongly influenced by the optical thickness.

Integrating the images for the Stokes parameters $I, Q, U$,
we estimate the total linear polarization from our spherical dust configuration
as one observes from the position perpendicular to the $z$-axis.
Because of the symmetry, it is the same along the $x$ and $y$-axes
\be
  P_{\rm l, \,90\degr} = \frac{F_{\rm sca, \,90\degr}^{\rm p}}
     {F_{\star}{\rm exp}\left[-\tau_{90\degr}^{\rm ext}\right]
     + F_{\rm sca,\,90\degr}}\,,   \label{Pol}
\ee
where $F_{\rm sca, \,90\degr}$ and $F_{\rm sca, \,90\degr}^{\rm p}$ are
the scattered and polarized scattered radiation from the configuration.
For the models presented in Fig.~\ref{optdepth}, the resulting polarization
is $P_{\rm l} = 4\%$, 9\%, and 8\%, if
$\tau_{\rm z} =$ 0.2, 1 and 2, correspondingly.
Excluding the stellar radiation (first term in the denominator in
Eq.~(\ref{Pol})), we can find the maximum degree of the observed
linear polarization
\be
  P_{\rm l, \,90\degr, \, max} = \frac{F_{\rm sca, \,90\degr}^{\rm p}}
     {F_{\rm sca,\,90\degr}}\,.   \label{P_m}
\ee
It is equal to
$P_{\rm l, \,90\degr, \, max} = 25\pm 1\,\%$, $15\pm 1\,\%$,
and $9.1\pm 1\,\%$, if
$\tau_{\rm z} =$ 0.2, 1 and 2, correspondingly.

\subsubsection{Dependence on dust grain parameters} \label{grpar}

\begin{figure*} 
\centering
 \resizebox{\hsize}{!}{\includegraphics{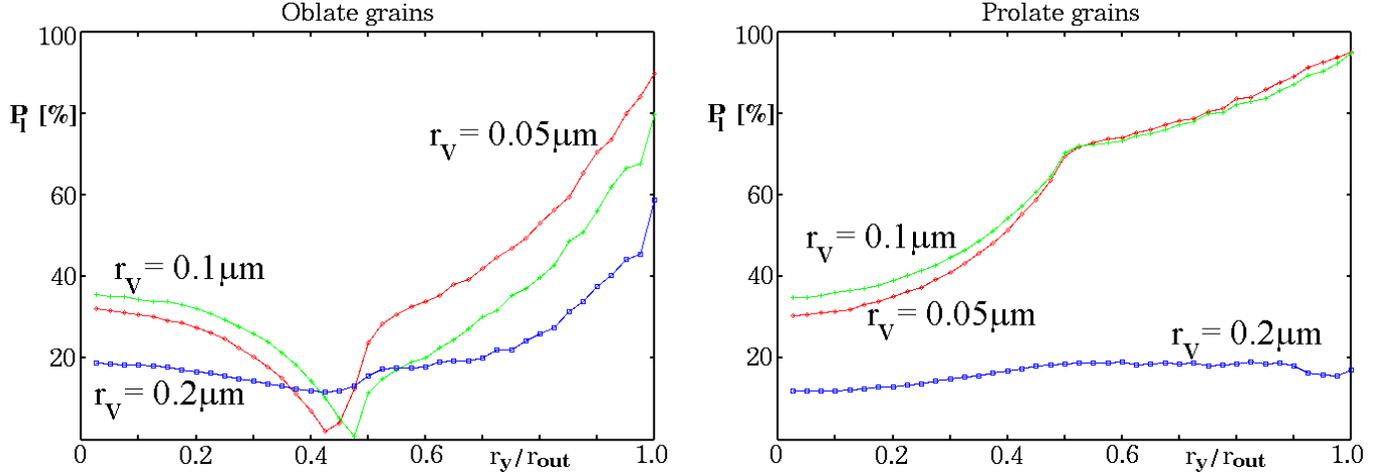}}
  \caption[]{
    The degree of linear polarization
    in dependence of the projected distance.
    The effect of the variations of the equivolume radius $r_{\rm V}$
    for oblate and prolate grains is illustrated 
    ($r_{\rm V}$ = 0.05$\,\mu$m, 0.1$\,\mu$m, 0.2$\,\mu$m).}
  \label{rv}
\end{figure*}
Figure~\ref{rv} shows the midplane polarization $P_{\rm l}(r_{\rm y}/r_{\rm out})$
for oblate and prolate grains of three different equivolume radii 
($r_{\rm V}$ = 0.05$\,\mu$m, 0.1$\,\mu$m, 0.2$\,\mu$m).
The behaviour of  polarization for particles with
$r_{\rm V}=0.05\,\mu{\rm m}$ and $0.1\,\mu{\rm m}$  is rather similar,
while the polarization produced by the largest particles
($r_{\rm V}=0.2\,\mu$m) is the smallest and is almost independent of
the projected radius. This feature reflects the fact of decreasing
polarizing efficiency of large particles
(Voshchinnikov et al.~\cite{vihf00}).
Because the manner of polarization for prolate grains
in the direction perpendicular to  the midplane is similar to that for
oblate grains in the midplane and vice versa
(see also Fig.~\ref{images}), the resulting polarization for particles of
both types is quite similar.
The net polarization --- excluding the direct, non-scattered stellar
radiation --- amounts to
$P_{\rm l, \,90\degr, \, max} =$
20\,\% (16\,\%),
25\,\% (20\,\%) and
11\,\% (10\,\%) for oblate (prolate) grains with equivolume radii
$r_{\rm V}=0.05\,\mu{\rm m}$,
$0.1\,\mu{\rm m}$ and
$0.2\,\mu{\rm m}$, respectively.

\begin{figure*} 
\centering
 \resizebox{\hsize}{!}{\includegraphics{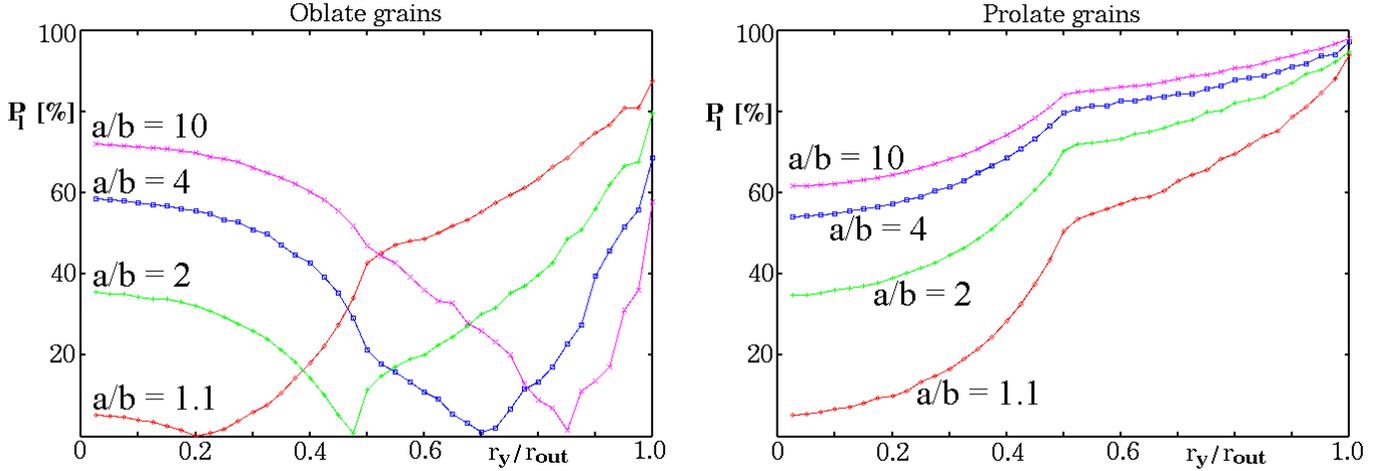}}
  \caption[]{
    The degree of linear polarization
    for the case of oblate and prolate grains
    in dependence of the projected distance.
    The effect of the variations of the grain shape (the aspect ratio)
    is illustrated ($a/b$ = 1.1, 2, 4, 10). }
  \label{a2b}
\end{figure*}
The distribution  of the linear polarization $P_{\rm l}(r_{\rm y}/r_{\rm out})$
in the midplane of the spherical shell is plotted in Fig.~\ref{a2b}
for particles with four different aspect ratios $a/b$
($a/b$ = 1.1, 2, 4, 10). The polarization minima in the midplane ($r_{\rm z}=0$)
appear for oblate grains only.
Their position  shifts to outer projected radii $r_{\rm y}/r_{\rm out}$
with increasing aspect ratio.  If  $a/b \rightarrow 1$,
the location of the polarization null point leads
to $r_{\rm y}/r_{\rm out} \rightarrow 0$.
This is consistent with the behaviour of the degree of linear polarization
for spherical grains where the null polarization can be found at
the centre for the scattering angle $\Theta=0\degr$.
The  resulting polarization for oblate and prolate grains
of different shape is given in Table~\ref{t1}.
Note that the polarization produced by oblate particles is larger in all cases.

\begin{table}[hb]
\centering
\caption[]{The net linear polarization from the spherical
shell containing  perfectly aligned
spheroidal grains ($m=1.7+0.03i$, $r_{\rm V} = 0.1\,{\rm \mu}$m)
as a function of the grains' axes ratio $a/b$.
The direct stellar radiation is excluded.}    \label{t1}
\begin{tabular}{ccc}
\hline
$a/b$  &  Oblate particles & Prolate particles  \\
\hline
  1.1  &    3.6\,\%   &   3.0\,\%  \\
  2    &     25\,\%   &    20\,\%  \\
  4    &     43\,\%   &    31\,\%  \\
 10    &     57\,\%   &    41\,\%  \\
\hline
\end{tabular}
\end{table}

\begin{figure*}
\centering
\resizebox{\hsize}{!}{\includegraphics{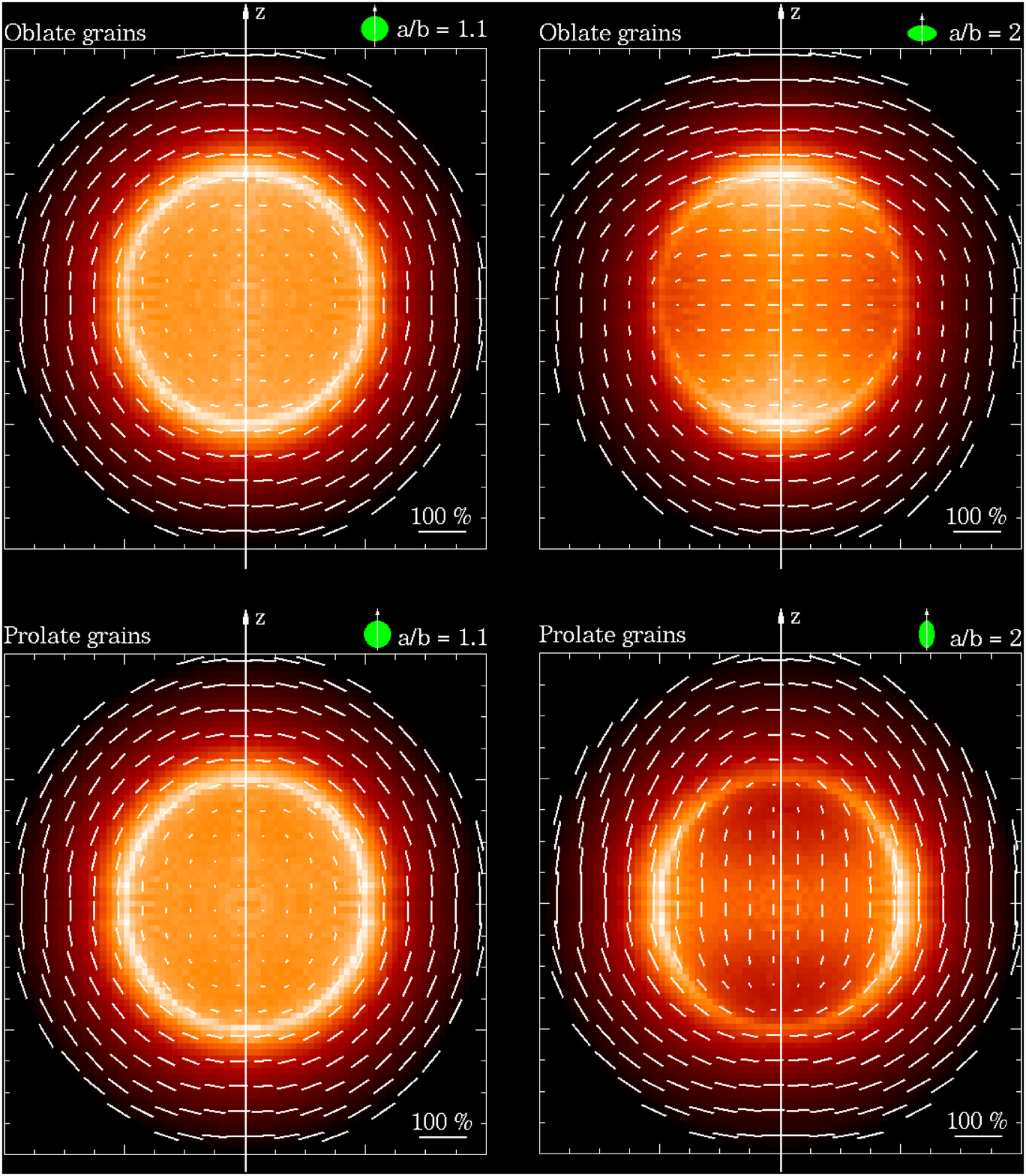}}
  \caption[]{
    Intensity maps with overlaid polarization pattern
    of a spherical shell containing perfectly aligned dust grains
    (for the description of the model configuration see Sect.~\ref{grpar}).
    Due to chosen shell and grain parameters,
    the brightest knots appear
    at the inner boundary of the shell.}
  \label{images}
\end{figure*}

The intensity maps of our configuration ($i=90^{\rm o}$) show an
additional effect. In Fig.~\ref{images}, the images
with overlaid polarization pattern, corresponding to the
intensity and linear polarization of scattered light,
are shown for oblate and prolate grains with the aspect ratios $a/b$=1.1 and 2.
In the case of nearly spherical grains ($a/b=1.1$), the ring-like structure
is clearly seen for oblate as well as for prolate grains.
Otherwise, in the case of $a/b=2$ the  intensity maxima appear
at the points of intersection with the $z$($y$)-axis
for oblate (prolate) grains. In both cases, the corresponding axis
is parallel to the minor ($b$) axis of grains.
Increasing the aspect ratio $a/b$, this effect is strengthened.
Note that the net polarization from the spherical configuration
arises in the inner parts of the shell because
the polarization pattern from the outer regions is centro-symmetric
and the polarization would be cancelled. The direction of the resulting 
net polarization is parallel to the major ($a$) axis of grains that is similar
to the case of dichroic polarization.

\subsubsection{Null points on polarization maps
as indicators of scattering by non-spherical grains} \label{null}

{\rm The model discussed in this paper is still too simple in order to make a
detailed comparison with observations. 
However, we will discuss qualitatively the behaviour of null points
produced by scattering on non-spherical grains.

Polarization null points were observed in a variety of objects
ranging from reflection nebula and circumstellar structures
to external galaxies
(e.g., Scarrott et al.~\cite{sdw89};  Scarrott et al.~\cite{srs90};
Gledhill \& Scarrott~\cite{gs89};  Asselin et al.~\cite{assetal96};
Chrysostomou et al.~\cite{chryetal96};
Hajjar \& Bastien~\cite{hb96}; Kastner \& Weintraub~\cite{kw96};
Wood \& Jones~\cite{wj97};
Gledhill et al.~\cite{getal01}).
They appear approximately symmetrically on either side of
the central illuminating source\footnote{It should be mentioned that
the measurable polarization --- especially near the position of polarization
null points --- sensitively depends on the spatial resolution
of the polarization maps. Therefore, it is more likely to find
the polarization minima but not real null points.}.
In {\it all} cases the origin of  null
points was attributed to the combined action of two polarization mechanisms,
where the first (initial or internal) mechanism produced linearly polarized radiation
whereas the second (external) mechanism led to the compensation
of this polarization.
Three possible models were qualitatively  discussed by
Scarrott et al.~(\cite{sdw89}):

(i) Polarization due to scattering by spherical grains and
its cancellation via dichroic extinction in an external medium consisting
of aligned non-spherical particles;

(ii) Polarization due to scattering by spherical grains
in an optically thick internal medium and
its cancellation after the next scattering by spherical grains
in an external optically thin medium;

(iii) Polarization of light from the central source due to an unspecified
mechanism and its compensation due to the
second scattering by spherical grains
in an external medium.

\noindent These models could be called ``hybrid models'' because
they require two steps to produce the observed polarization null points.
The model (i) is rather naturally applied to the
modelling of polarization maps of galaxies
(Wood \& Jones~\cite{wj97}).
The multiple scattering model (ii) was used by
Bastien \& M\'enard~(\cite{bm88}) and Fischer et al.~(\cite{fhy94})
for interpretation of polarization maps of young
stellar objects. In this case, the centrosymmetric polarization
pattern is obtained in outer optically thin parts of the shell,
whereas the aligned polarization vectors appear in the dense inner parts
of this shell.
The polarized source model (iii) was discussed by
Notni~(\cite{n85}) and Gledhill~(\cite{g91}).
Using this model, it is possible to explain the polarization
pattern observed in several bipolar and cometary nebulae (e.g., Gledhill \cite{g91}),
but as it was shown by Clark et al.~(\cite{cetal00}) that
the models with spherical grains fail to
produce  even the 5\,\% circular polarization found
in the reflection nebulosity around the Herbig Ae/Be star R~CrA.
In this case, the authors suggested scattering by aligned
non-spherical grains as the operating mechanism.

In contrast to the hybrid models,
different polarimetric effects can arise as a result
of light scattering by aligned non-spherical particles.
In our simple model, the location of polarization null
points depends on the particle shape, size and
the inner to outer radius ratio (see
Figs.~\ref{shell2sphere}, \ref{rv} and \ref{a2b}).
The wavelength dependence of the location of
null points is not very strong and their position only slightly approaches the 
central (stellar) position when going from the V to K band
as it follows from Fig.~\ref{nuu}\footnote{In the case
of polarization null points which have been observed at the
location of the outer boundary of circumstellar disks around young
stellar objects (transition from the optically thick disk to the optically thin shell),
the same behaviour is expected since this boundary is slightly
shifted inward with increasing wavelength.}.
\begin{figure}[htb]
\centering\resizebox{\hsize}{!}{\includegraphics{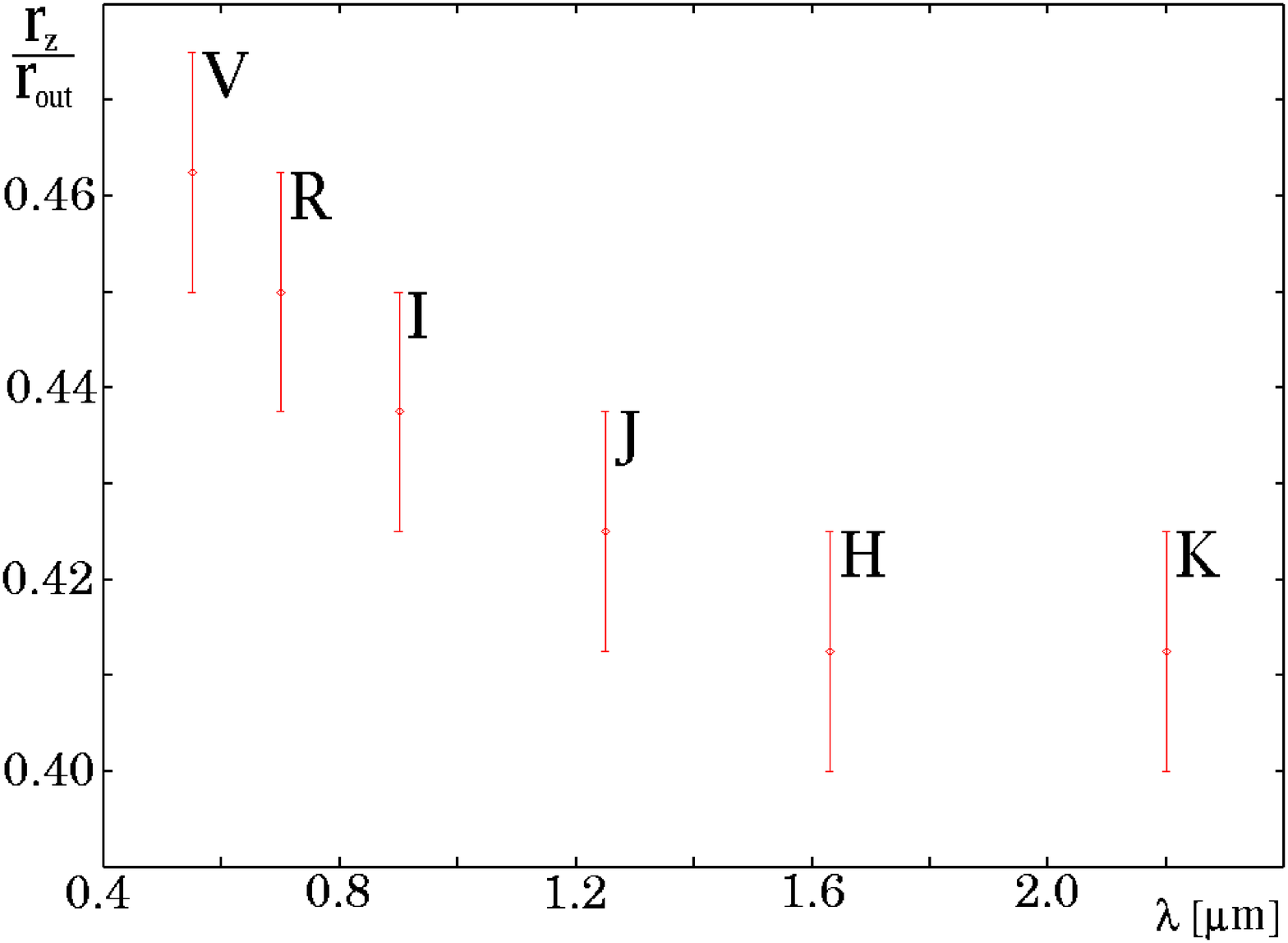}}
  \caption[]{
    Projected positions of  null points on polarization maps
    in dependence of wavelength and their statistical errors
    for the case prolate grains with aspect ratio $a/b = 2$
    (the similar model is shown in Fig.~\ref{images},  right lower panel).
    The calculations were performed for effective wavelengths
    of Johnson photometric system.}
  \label{nuu}
\end{figure}
A similar conclusion can be reached from
the observed polarization maps of the variable
Hubble's nebula NGC~2261 (Scarrott et al.~\cite{sdw89})
and the infrared reflection nebula  GSS~30
(Chrysostomou et al.~\cite{chryetal96}).
The results of Asselin et al.~(\cite{assetal96}) for V376~Cas
show that the position of the null points is also shifted as a function of
the binned box size used in averaging the polarization.
However, it should be noted that near the null points the observed
signal is usually very small and the errors are large.

Finally, we can indicate that the scattering by almost spherical
grains is  marked by a  ring-like structure in the intensity maps
whereas the null point are located inside this structure rather
close to the stellar position (Fig.~\ref{images}, left side).
In the case of aligned non-spherical particles (Fig.~\ref{images}, right side),
the null points at the inner boundary of the spherical shell
are accompanied by symmetric arc-like structures in the intensity
maps. There is some similarity of such maps with those of
small reflection 
nebulae published by Gledhill et al.~(\cite{getal01}):
the ``spherical case'' is seen in the source IRAS~19114+0002 while
the ``non-spherical case'' occurs in the maps of IRAS~22223+4327,
IRAS~19500-1709 and IRAS~17106-3046.


We should emphasize that {\it in the presented model
the origin of null points is caused by light scattering
by non-spherical particles}, but is not related to the
cancellation of polarization as in hybrid models.
}

\section{Conclusions}\label{concl}

The paper contains the first solution to the RT problem
including multiple light scattering
in dust configurations with aligned non-spherical dust grains.
Our main aim was  to discuss the effects of light scattering by spheroidal
dust grains in a simple (spherical) dust configuration, containing
perfectly aligned particles.

The most remarkable features of the simulated
linear polarization maps are the polarization null points where the reversal of
polarization occurs. They appear when the grain alignment
axis is perpendicular to the line of sight.
Symmetrically to the polarization null points,
we found maxima in the intensity maps.

In contrast to spherical grains, even single scattering by spheroidal grains
may cause circular polarization.
The maps of circular polarization have a sector-like
structure with polarization maxima at the ends of a line inclined
to the grain alignment axis by $\pm 45\degr$.

{\rm Based on our theoretical investigations of light scattering,
dichroic absorption, and (re)emission by spheroidal grains presented
in Sect.~\ref{basic-all}, the next steps of the numerical radiative transfer 
simulations will include the consideration of light scattering by
partly aligned rotating particles and the calculation of the polarized
re-emission of grains in different dust configurations.
This will immediately lead to an improved
interpretation and therefore better understanding
of mid-infrared and submillimeter polarization 
where re-emission and dichroic absorption by partly aligned non-spherical
dust grains are the main polarization mechanisms.
Thus, it provides a basis for investigations of the structure
of the magnetic fields in these objects which are
in most cases assumed to cause grain alignment.
}

\begin{acknowledgements}
We appreciate the discussions of light scattering by
non-spherical particles with Victor Farafonov and Bernhard Michel
and wish to thank M.\ Matsumura for providing his results for comparison.
We are grateful to the referee for  constructive comments.
This research was supported by the DFG grant Ste 605/10 within the program
``Physics of star formation'' and by grants of the Volkswagen Foundation
and INTAS (Open Call 99/652).
\end{acknowledgements}

{}
\end{document}